\documentclass[10pt]{iopart}
\usepackage{iopams}  %Uncomment next line if AMS fonts required.
\usepackage{graphicx}
\usepackage{subfigure}
\usepackage{booktabs}
\usepackage{cite}    %Package for condensed References.
\usepackage{xcolor}
\usepackage{hyperref}
\hypersetup{colorlinks,urlcolor=blue,linkcolor=blue,citecolor=blue,filecolor=blue,urlcolor=blue}
\usepackage{array,longtable}

\usepackage{etoolbox}
\makeatletter
\newif\ifLT@nocaption
\preto\longtable{\LT@nocaptiontrue}
\appto\endlongtable{%
  \ifLT@nocaption
    \addtocounter{table}{\m@ne}%
  \fi}
\preto\LT@caption{%
  \noalign{\global\LT@nocaptionfalse}}
\makeatother

\begin{document}

\title {On the distribution of spontaneous potentials intervals in nervous transmission}
%%%%%%%%%%%%%%%%%
\author{A J da Silva$^1$,  S Floquet$^2$ and R F Lima$^{3}$ }
\address{$^1$ Instituto de Humanidades, Artes e Ci\^encias, Universidade Federal do Sul da Bahia, $45613-204$, Itabuna, Bahia.  Brazil}
\address{$^2$ Colegiado de Engenharia Civil, Universidade Federal do Vale do S\~ao Francisco, $48902-300$, Juazeiro, Bahia, Brazil}
\address{$^{3}$ Departamento de Fisiologia e Farmacologia, Faculdade de Medicina, Universidade Federal do Cear\'a, $60430-270$, Fortaleza, Cear\'a, Brazil}

\ead{adjesbr@ufsb.edu.br,adjesbr@gmail.com}

\begin{abstract}

One of the main challenges in Biophysics teaching consists on how to motivate students to appreciate the beauty of theoretical formulations. This is crucial when the system modeling requires numerical calculations to achieve realistic results. In this sense, due to the massive use of software, students often become a mere users of computational tools without capturing the essence of formulation and further problem solution. It is, therefore, necessary for instructors to find innovating ways, allowing students developing of their ability to deal with mathematical modelling. To address this issue one can highlight the use of Benford's law, thanks to its simple formulation, easy computational implementation and wide possibility for applications. Indeed, this law enables students to carry out their own data analysis with use of free software packages. This law is among the several power or scaling laws found in biological systems. However, to the best of our knowledge, this law has not been contemplated in Cell Biophysics yet. Beyond its vast applications in many fields, neuromuscular junction represents a remarkable substrate for learning and teaching of complex system. Thus, in this work, we applied both classical and a generalized form of Benford's Law, to examine if electrophysiological data recorded from neuromuscular junction conforms the anomalous number law. The results indicated that nerve-muscle communications conform the generalized Benford's law better than the seminal formulation. From our electrophysiological measurements a biological scenario is used to interpret the theoretical analysis. 

\end{abstract}

% Uncomment for keywords
\vspace{2pc}
\noindent{\it Keywords}: anomalous numbers, biophysics, nervous transmission, Benford's law
% For two-column output uncomment the next line and choose [10pt] rather than [12pt] in the \documentclass declaration
%\ioptwocol
%\submitto{\EJP}% Uncomment for Submitted to journal title message
\maketitle

\normalsize

\section{Introduction}

The strong interdisciplinary character of Neuroscience and Biophysics opens wide possibilities for learning several quantitative aspects of biological systems. However, due to several reasons, teaching Biophysics modeling still represents a challenge \cite{islami}. Among factors one can mention the inappropriate introduction of the phenomenon prior the theoretical model development and the lack of motivating didactic sequence. In order to overcome this challenges, one can highlight a counterintuitive law, represented by a logarithmic distribution function, also known as First Digit Law or "Law of anomalous numbers". It was firstly discovered by polymath Simon Newcomb in $1881$, but only after $57$ years the seminal observation was revisited and popularized thanks to the physicist Frank Benford, analyzing different amount of data. Since then, the scientific community consolidated this peculiar numerical pattern as Benford's Law (BL) \cite{newcomb,benford}. As it is shown in table \ref{tab1} this law predicts the proportional frequency of each leading digit.

\begin{table}[!h]
\centering\caption{Theoretical logarithmic law (the number zero can not be the first significant digit).}\label{tab1}
\begin{indented}
\item[]
\begin{tabular}{@{}lccccccccc@{}} 
\br
%\toprule
 First digit &  $1$   & $2$   & $3$   & $4$  & $5$  & $6$  & $7$  & $8$  & $9$ \\ \mr
 Frequency(\%)  &  $30.1$& $17.6$& $12.5$& $9.7$& $7.9$& $6.7$& $5.8$& $5.1$& $4.6$ \\
%\bottomrule        
\br
\end{tabular}
\end{indented}
\end{table}

According BL, the probabilities for occurrences of leading digits is inferred by: 

\begin{eqnarray}
 P_{i} & = & \log_{10}\left( \frac{i+1}{i}  \right), \qquad  i \in \left\lbrace 1,2,3,...,9 \right\rbrace. \label{benford}
\end{eqnarray}

Beyond the seminal heuristic formulation, sophisticated mathematical descriptions for explaining the significant first digit existence has been proposed. In this framework, many important properties were systematically discovered. For instance, Pinkham argued that if BL obeys some universal distribution, then this law should be scale invariant to the units chosen \cite{pinkham}. In these scheme, if data follows BL, they must exhibit a base invariance profile that is expressed in any other base \cite{hill95}. In fact, Hill contributed with a rigorous statistical pillar, inserting the law as a branch of modern probability theory, showing that invariance uniquely implies BL \cite{hill95}. To quantify how data tends to the frequencies predicted by BL, many explanations arose including the resurgence of the classical argument given by Newcomb. From Benford findings one can elaborate a crucial question: Is it BL ubiquitous in nature? Investigations have shown distinct phenomena, which apparently do not obey the law despite displaying the typically asymmetric distribution. Examples corroborating a non ubiquity argument include seismic activity, distribution of the Discrete Cosine Transform and quantized JPEG coefficients and cognition experiments \cite{pietronero,fu2007,gauvrit2009}. Naturally, to expand the BL validity is necessary extending the pioneering formulation. Among alternative models proposed, one can highlight the generalization introduced by Pietronero \textit{et al.} as written below:

\begin{eqnarray}
 P(n) & = & \int_{n}^{n+1} N^{-\alpha} dN \label{eq-pn}
\end{eqnarray}
that is the differential equation:
\begin{eqnarray}
 \frac{dP(N)}{dN} & = & N^{-\alpha}
\end{eqnarray}

Solving (3) results in a $\alpha$-logarithm one obtain the solution:

\begin{eqnarray}
 P_{\alpha}(n) & = & \frac{1}{1-\alpha}\left[ \left(n+1\right)^{(1-\alpha)} - n^{(1-\alpha)}\right] \\
   & = & n^{(1-\alpha)} \ \ln_{\alpha}\left( \frac{n+1}{n} \right)
\end{eqnarray}

\begin{figure}[!htb]
\begin{center}
\includegraphics[width=12cm]{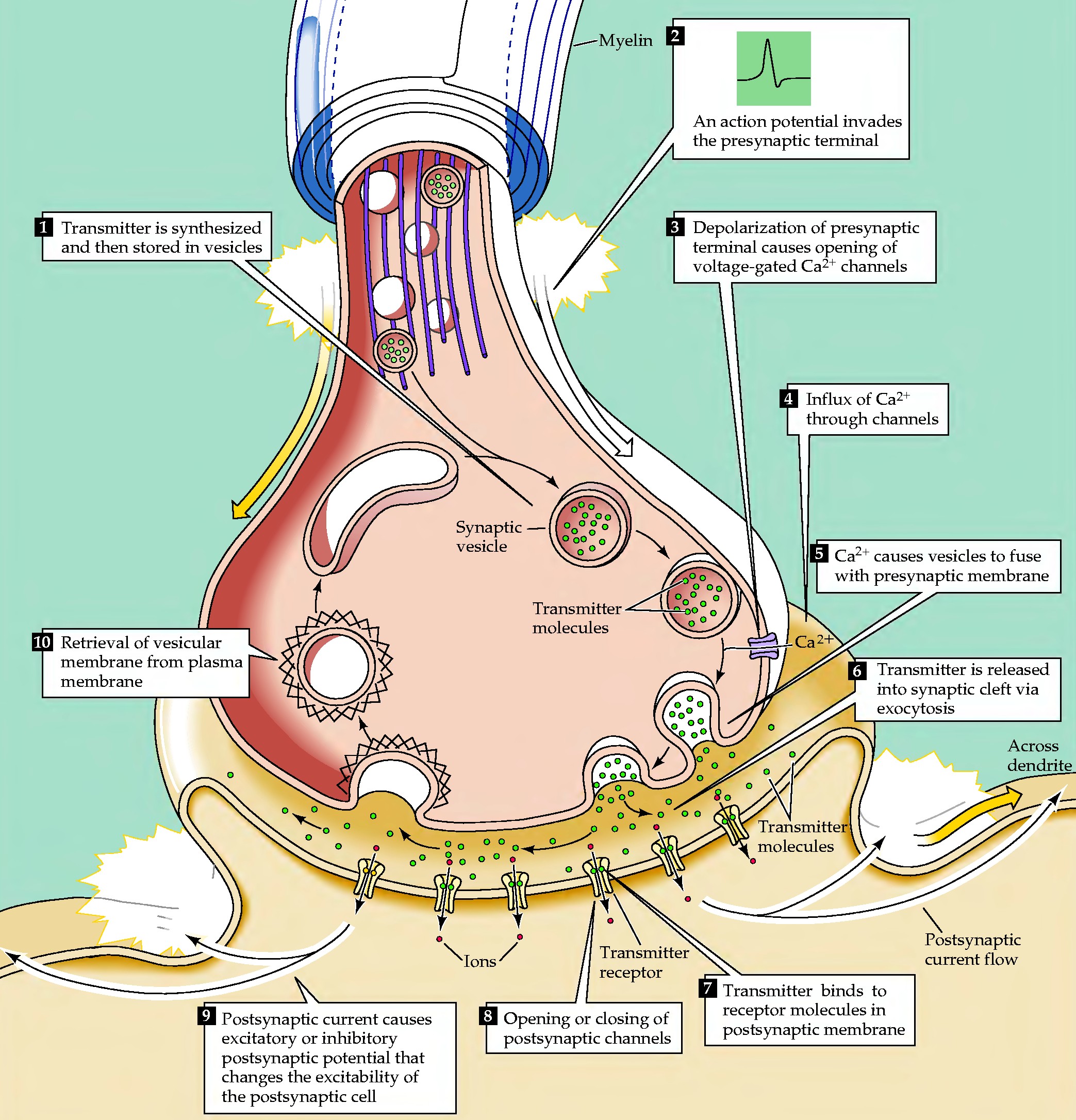} 
\end{center}
\caption{\label{fig1} Sequence of events observed in a typical synaptic transmission. Figure used by permission of author \cite{purves}.}
\end{figure}

 According to (5), defined as generalized BL (gBL), more frequent first digits than expected by BL implies in $\alpha>1$, while $\alpha<1$ means a first digit frequency below of the predicted percentage. As expected, when $\alpha=1$ the classical law is recovered. BL has been observed in different kinds of statistical data found in physical constants, number of cells in colonies of the cyanobacterium, alpha decay half-lives and fraud-detection \cite{burke,costas,buck,nigrini}. In Physiology, BL applications involve dynamical transitions in cardiac models and brain electrical activity \cite{sinha2015,kreuzer2014}. Nevertheless, BL and gBL remaining to be verified in another genuine complex system: the synaptic terminal. Cell interaction is accomplished by a complicated network of molecular signaling whose all details are not fully understood. In particular, information processing in the central nervous system (CNS) is mainly accomplished by a specialized structure called chemical synapses. Among many physiological functions, these zones are involved in nerve-muscle communication. Synaptic transmission is mediated by one or more substances called neurotransmitters \cite{sudhof}. The transmission is basically accomplished in the following steps: An action potential (AP) arrives in the nerve ending, promoting the depolarization of specialized proteins known as calcium channels voltage-dependent; Calcium ions flow through the channel diffusing inside the nervous terminal; into the nerve hey trigger the fusion of vesicle placed on the active zone (AZ) resulting in the releasing of neutransmitters, transmitting the impulse  transmission to the neighbor cell. Figure \ref{fig1} brings a more detailed view of neurotransmission steps \cite{purves}.

%\begin{figure}[!htb]
%\begin{center}
%\includegraphics[width=10cm]{fig1.eps}
%\end{center}
%\caption{Miniature end-plate potentials taken from a data set.}
%\label{fig1}
%\end{figure}

The neuromuscular junction (NMJ) is responsible for communicating electrical impulses from the motor neuron to the skeletal muscle to signal contraction \cite{sanes}. The terminal formed by the NMJ constitute a standard example of the well-studied chemical synapse. The facility on tissue extraction and stereotyped electrical response of nervous activity constitute the main advantages to adopt NMJ in both teaching and research. Bernard Katz led most of the pioneering work on the biophysics related to the neuromuscular transmission \cite{katz1}. Using electrophysiological recordings Katz and colleagues discovered a very stereotypical spontaneous electrical activity, which plays role on the nervous terminal, called miniature end-plate potential (MEPP) \cite{katz2}. These electrical signals were attributed to single vesicle fusing with the membrane terminal. Since then, NMJ established itself as the classical pillar in synaptic transmission investigation, being crucial for the understanding of mechanisms involved in neurotransmitter release. In this context, electrophysiological recordings revealed that MEPPs are no longer constant in size or temporal distribution, associating MEPPs generation as governed by Gaussian and Poisson model, respectively \cite{bennett}. Katz, himself, curiously attempted to point out weaknesses in the Poissonian predictions. In fact, further introduction of a more refined experimental design, combined with powerful computational approach, exposed the nonrandom pattern embedded in several NMJ preparations \cite{washio,lowen,takeda}. For example, these investigations showed the fractal behavior in quantal release. Additionally, morphological studies provided a detailed scenario supporting a non-haphazard mechanism, credited to the complex physical structure of the terminal itself \cite{rizolli}. These structural analysis of NMJ revealed the presence of thousands of crowding vesicles sharing a very confined volume into the nerve terminal. Altogether, the arguments given above clearly configures the NMJ as a suitable preparation for several applications into education and research of Biophysics. 

\section{Materials and Methods}
 
\subsection{Experimental data acquisition}

\begin{figure}[!htb]
\begin{center}
\includegraphics[width=8 cm]{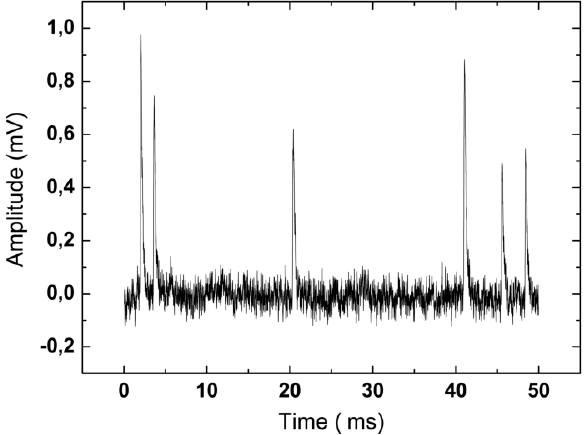}
\end{center}
\caption{A representative electrophysiological recording portion showing MEPPs.}
\label{fig2}
\end{figure}

The experimental procedures here adopted were approved by the Animal Research Committee (CETEA - UFMG, protocol 073/03), and described more precisely in other study \cite{caboco}. Briefly speaking, adult mice were euthanized by cervical dislocation, diaphragms were extracted and inserted into a physiological artificial fluid (Ringer solution) containing the following concentrations (mM): NaCl $(137)$, NaHCO$_{3}$ $(26)$, KCl $(5)$, NaH$_{2}$PO$_{4}$ $(1.2)$, glucose $(10)$, CaCl$_{2}$ $(2.4)$, MgCl$_{2}$ $(1.3)$, and pH adjusted to $7.4$ after gassing with $95\%   O_{2}$ - 5\% CO$_{2}$. It is valid mentioning that $[Ca^{2+}]_{o}$ physiological concentration of mammalians is about $1.8$ mM. The tissues were left bathing in the solution $30$ minutes before the electrophysiological recordings began, to minimize the mechanic trauma of their extraction. Next, tissues were gently transferred to a recording chamber continuously bathed with artificial solution at $T = 24 \pm 1^\circ{}C$. Standard intracellular recording technique was performed to monitor the frequency of spontaneous MEPP by inserting a micropipette at the chosen muscle fiber. The electrophysiological acquisition was performed using borosilicate glass microelectrodes with resistances of $8-15$ M$\Omega$ when filled with $3$ M KCl. Strathclyde Electrophysiology Software (John Dempster, University of Strathclyde), R packages \cite{rsoftware}  and Origin (OriginLab, Northampton, MA) were employed for electrophysiological acquisition and data analysis. Please see the source code, used to perform the simulations, available into Supplementary Material section. For a rigorous analysis, $33$ recordings, taken from different fibers with at least $1000$ MEPP intervals, were used. Figure \ref{fig2} shows a representative segment from an electrophysiological recording. 

In living organisms, ions are responsible for several functions as changing solubility and melting temperature modulation of proteins. For example, potassium ions are responsible for resting membrane potential, sodium triggers the action potential generation, while calcium is crucial for the exocytosis and synaptic transmission in the ending terminal \cite{rusakov2006}. In this sense, MEPP frequency can also be manipulated by modification of the ionic concentration in Ringer solution. For this reason, the impact of $[Ca^{2+}]_{o}$ on the MEPP rate was verified assuming successive concentration (mM): $0.6$, $1.2$, $1.8$, $2.4$ and $4.8$. As a consequence, this experimental protocol enables indirectly verifying how robust is BL or gBL as function of vesicular fusion dynamics. It is also well-accepted the relationship between the membrane cellular potential and MEPP frequency. The discharge rate increases for more depolarized or positive potentials. This empirical environment may be achieved incrementing the extracellular content of potassium ($[K^{+}]_{o}$). Moreover, there are other argumentations supporting $[K^{+}]_{o}$ manipulation:(a) modulation in $[K^{+}]_{o}$ closely reproduces a physiological stimulation \cite{grohovaz}; (b) its effects on the membrane potential in muscle preparations is well understood \cite{adrian1956}; (c) it allows investigating BL and gBL conformity in a hyperexcitable neuronal state, by accelerating the MEPP frequency. Thus, application of BL and gBL, in conjunction with $[Ca^{2+}]_{o}$ and $[K^{+}]_{o}$ manipulation, constitute a relevant strategy to understand the scale-invariance in time series collected from electrophysiological recordings. Table \ref{tab2} brings the identification for each data for different concentrations to be used in the present work. 

\begin{table} 
\caption{Ions concentrations with their respective number of MEPP and data label.}\label{tab2}
\begin{indented}
\item[]\begin{tabular}{@{}ccc}
\br
Ion & Concentration (mM) & Data \\

\mr 
 $Ca^{2+}_{o}$ &  $0.6$  &   $25$, $26$, $27$ and $28$ \\
 $Ca^{2+}_{o}$ &  $1.2$  &   $19$, $20$, $29$ and $30$   \\
 $Ca^{2+}_{o}$ &  $1.8$  &   $6$, $13 $, $15 $, $16 $, $21$ and $22 $  \\
 $Ca^{2+}_{o}$ &  $2.4$  &   $1 $, $10 $, $11 $, $12 $, $14 $, $17$ and $18$ \\
 $Ca^{2+}_{o}$ &  $4.8$  &   $2 $, $3 $, $4 $, $5$, $7$, $8$ and $9$ \\
 $K^{+}_{o}$   &  $25.0 $  & $23 $, $24$,  $31$, $32$ and $33$ \\
\br
\end{tabular}
\end{indented}
\end{table}

\section{Results}

In this work, we address if the interval between MEPPs may conform BL or gBL. In more specific terms it will be assessed the following issues:
 
	a) Are the BL or gBL observed at physiological concentrations? 
	
	b) Is the conformity independent of the ionic concentration? 
	
	c) Is the conformity degree modulated by $[Ca^{2+}]_{o}$ and $[K^{+}]_{o}$ modifications?
	
	d) What are the biophysical implications? 
	 
To pursue these questions, choosing reliable statistical methods is an important requirement in order to achieve reliable results. In this sense, many authors assume the $\chi^{2}$ test for verifying BL conformity, although this test manifests the "excess of power" problem. Thus, to circumvent the "excess of power" effect, one can use the mean absolute deviation (MAD) or sum of squares difference (SSD) as suggested for Nigrini and Kossovsky, respectively \cite{nigrini,kossovsky}. According to MAD results between $0.000-0.006$ means "close conformity", while $0.006-0.012$ represents an "acceptable conformity" and $0.012-0.015$ configures a "marginal conformity". Values higher than $0.015$ are considered "non-conforming". In mathematical form MAD is defined as:

\begin{eqnarray}
 MAD & = & \frac{ \displaystyle \sum_{i=1}^{n} | AP_{i} - EP_{i}  |}{n}
\end{eqnarray}
where AP is the actual proportion and EP is the expect proportion.

In the same way, for SSD between $0-2$, $2-25$, $25-100$ represents "perfect conformity", "acceptable conformity" and "marginal conformity", respectively. Values above $100$ are considered "non-conforming". The SSD is calculated using the expression:

\begin{eqnarray}
 SSD & = & \displaystyle \sum_{i=1}^{n} \left( AP_{i} - EP_{i} \right)^{2}\times 10^{4}
\end{eqnarray}
Once again, AP is the actual proportion and EP is the expected proportion.

\begin{figure}[!htb]
\begin{center} \label{fig3} 
\includegraphics[width=6.4cm]{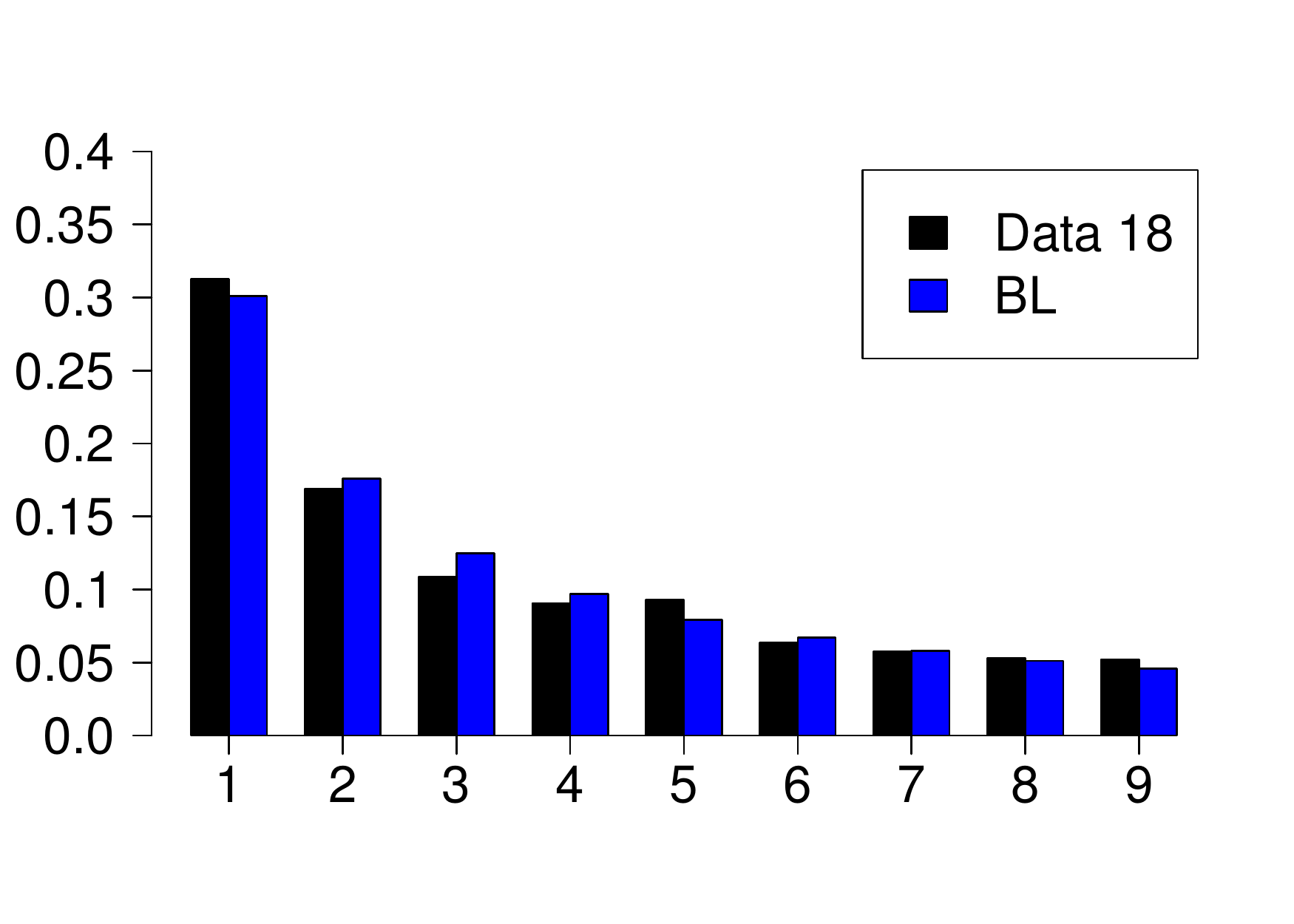} %\includegraphics[width=6.4cm]{bl-18.eps} 
\includegraphics[width=6.4cm]{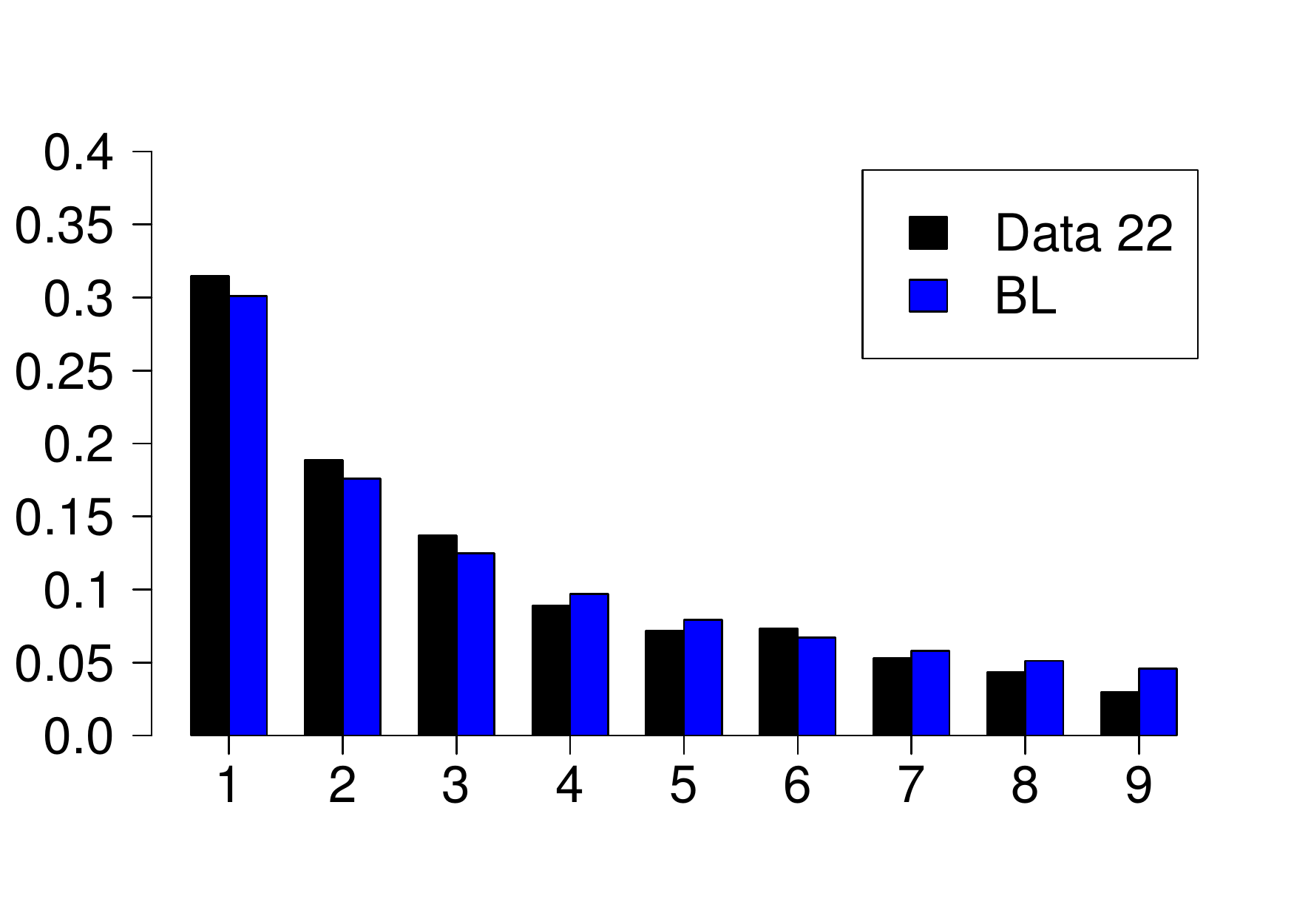} 

(a) \hspace{6.0cm} (b)

\includegraphics[width=6.4cm]{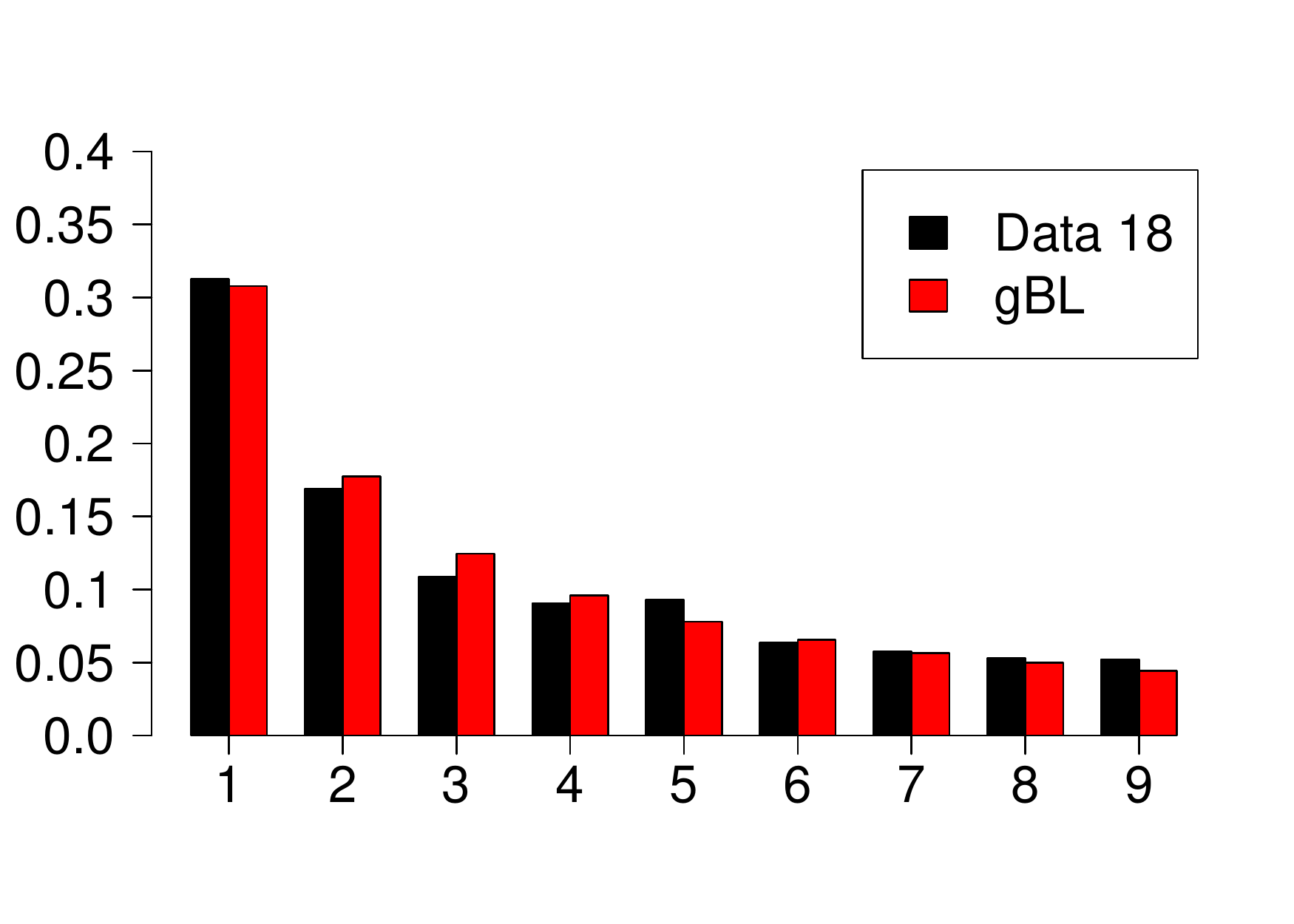} 
\includegraphics[width=6.4cm]{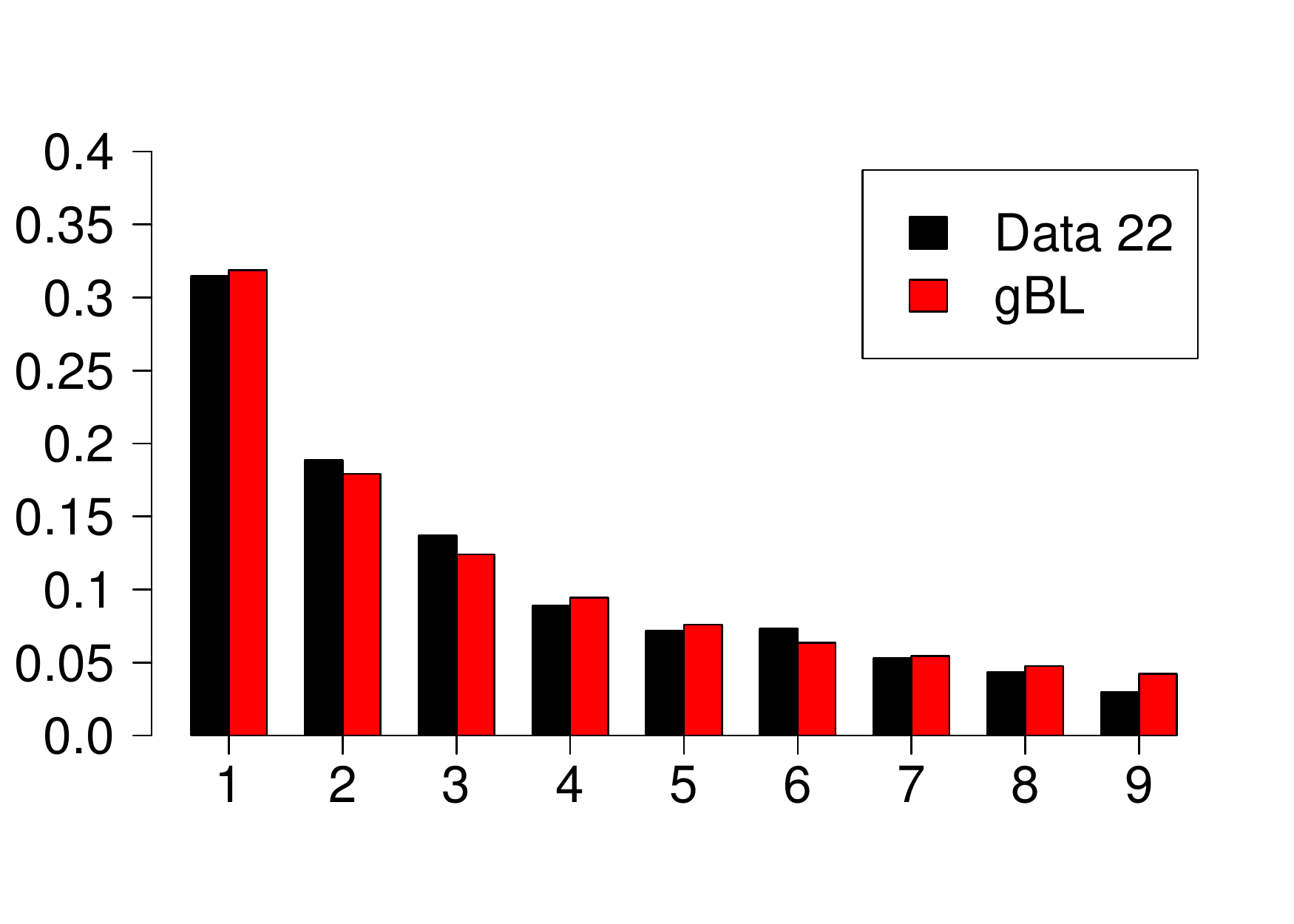} 

(c) \hspace{6.0cm} (d)

\end{center}

\caption{Representative examples showing two data set for $[Ca^{2+}]_{o}$ = $1.8$ mM concentrations using BL and gBL.}
\end{figure}

\begin{figure}[!htb] 
\begin{center} 
\includegraphics[width=6.3cm]{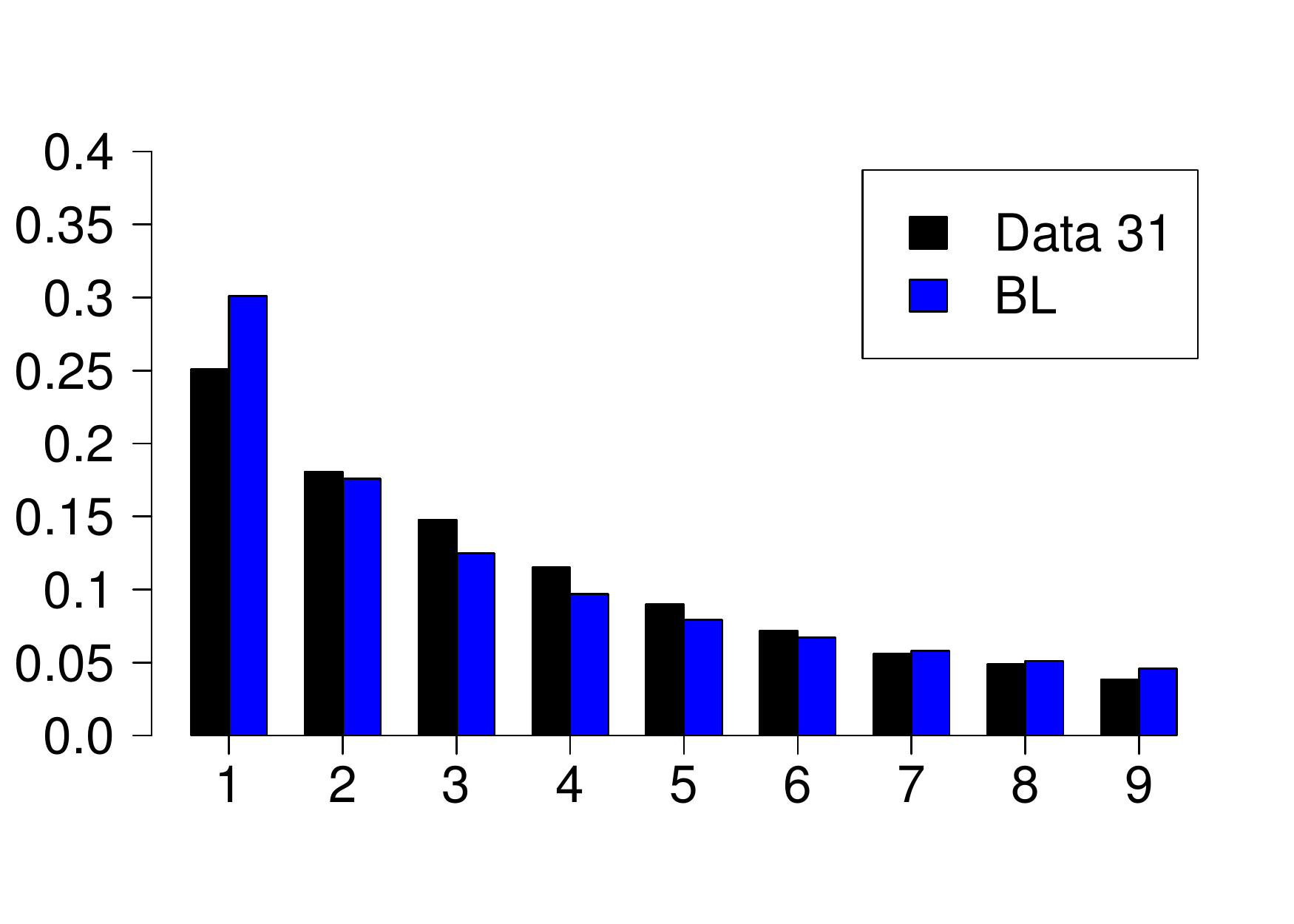} 
\includegraphics[width=6.3cm]{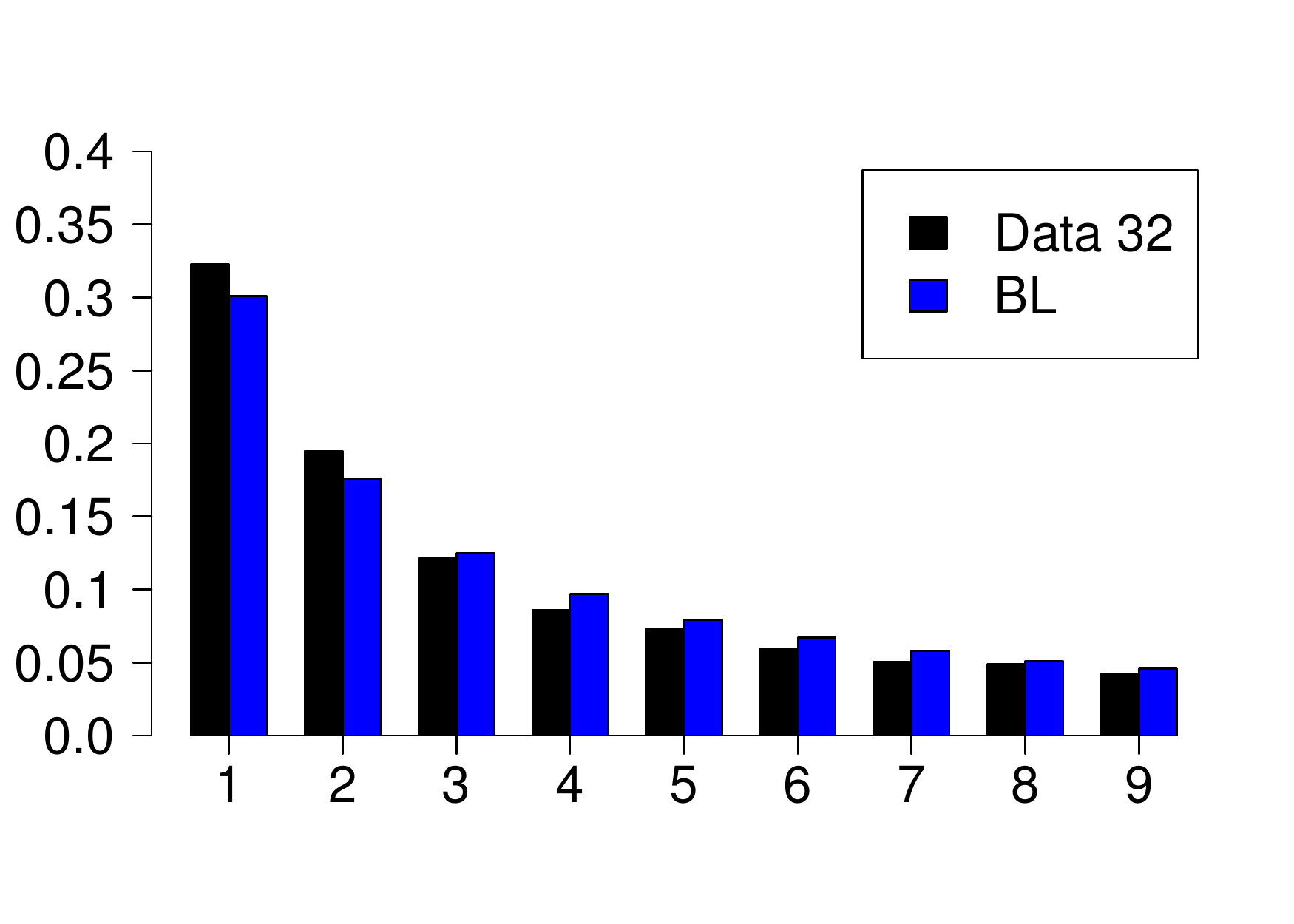} 

(a) \hspace{6.0cm} (b)

\includegraphics[width=6.3cm]{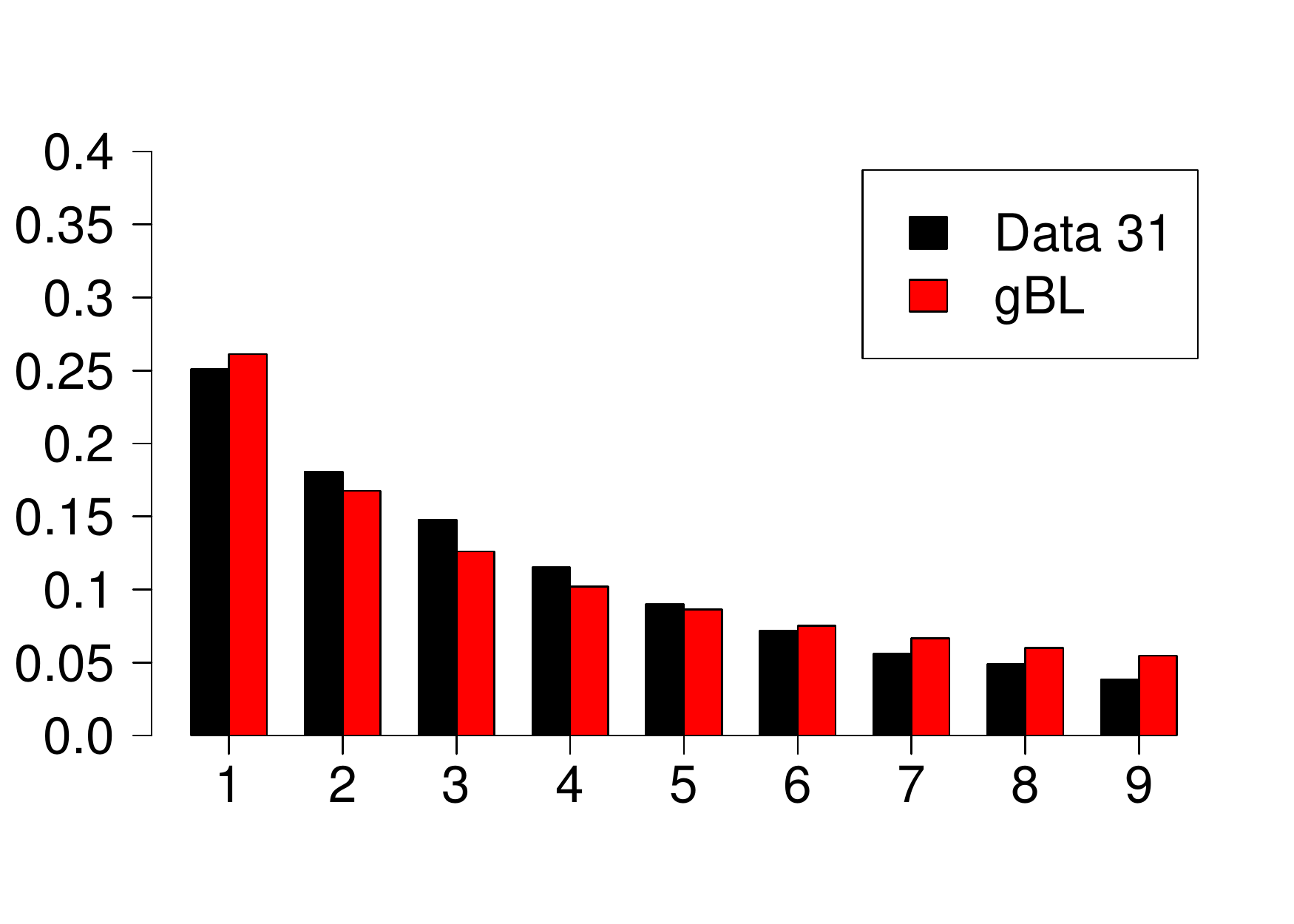} 
\includegraphics[width=6.3cm]{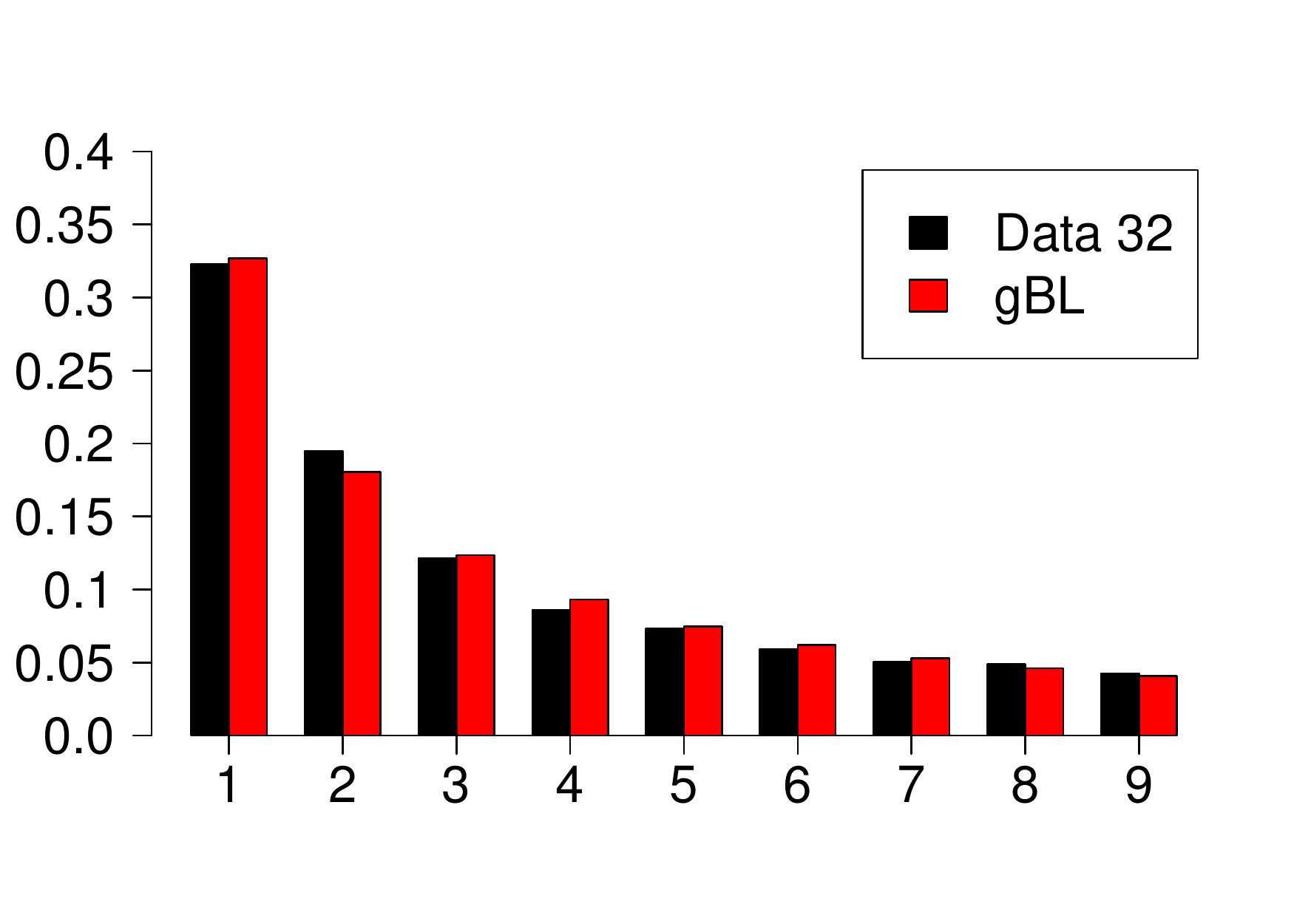} 

(c) \hspace{6.0cm} (d)
\end{center}

\caption{Representative figures showing results for high $[K^{+}]_{o}$. Top: MAD and SSD calculations revealed that both data $31$ and $32$ conforms BL. Bottom: Application of gBL, both data confirmed conformity in respect of SSD and MAD analysis.} \label{fig4}
\end{figure}

In this framework, using MAD and SSD, Slepkov \textit{et. al} uncovered BL conformity in answers of every end-of-chapter question in introductory physics and chemistry textbooks \cite{slepkov2015}. Using the same calculations, but including the gBL calculation, both statistical methods were here used to uncover a possible conformity in MEPP intervals. The results are summarized in tables \ref{tab3} and \ref{tab4}. Notably, SSD exhibited conformity in all cases analyzed with gBL, while conformity was computed in thirty-one cases when BL was used. Additionally, MAD revealed non-conformity in nine and five data for BL and gBL calculations, respectively. In particular, employing gBl to study physiological $[Ca^{2+}]_{o}$ recordings, a full conformity was documented in all data, while in only one case BL did not show conformity (data $13$). Figure \ref{fig3} shows the results for two electrophysiological recordings ($18$ and $22$), at physiological $[Ca^{2+}]_{o}$, where a conformity was satisfactorily obtained. Curiously, two data for high $[K^{+}]_{o}$ ($23$ and $24$) did not conform BL, regardless the statistical methods. In contrast, SSD confirmed that at high $[K^{+}]_{o}$ gBL is more robust to adjust these experimental results (figure \ref{fig4}). This situation was similarly extracted by Pietronero \textit{et al.} in their investigation using earthquake magnitude catalog. The table \ref{tab5} gives a concise tabulation of the present findings. 
 
\begin{table}[!htb]
\caption{A more synthetic presentation of the statistical analysis. }\label{tab5} 
\begin{indented}
\item[]\begin{tabular}{@{}ccc}
\br
Law & \multicolumn{2}{c}{Conformity} \\
\mr 
 BL            & \hspace{1.0cm} Yes \hspace{1.0cm} & \hspace{1.0cm}  No \hspace{1.0cm} \\
 MAD                     &   24  &   09  \\
 SSD                     &   31  &   02  \\ \mr 
 gBL &  Yes  &   No  \\
 MAD                     &  28   &   05  \\
 SSD                     &  33   &   0   \\
\br
\end{tabular}
\end{indented}
\end{table}

%\newpage

\section{Discussion}
   
	 Neurophysiology is an interdisciplinary area, propitious for the understanding of several problems in complex systems. In conjunction with available repository resources, instructors have an excellent opportunity to use novel empirical data, building a didactic sequence in modern topics of Biophysics \cite{halliday2010}. When successful, this strategy opens an plenty room for interesting applications in teaching and research \cite{falces2015,falces2013}. In this framework, the NMJ adoption, combined with the simplicity of BL and gBL analysis, certainly provide a creative form to explore new features of the probabilistic nature of biological systems \cite{branco}. Although the regulatory releasing mechanisms require a network of molecular cascade, exocytosis may be exacerbated when $[Ca^{2+}_{o}]$ and $[K^{+}]_{o}$ increases. According to the present results, both $[Ca^{2+}_{o}]$ and $[K^{+}]_{o}$ manipulation did not show any evident modulation with the conformity degree. In other words, different concentrations did not reveal a drastic deviation from the predicted anomalous numbers behavior. Thus, the results here presented allow to establish a validity of BL and gBL at the cellular level. From these observations one can conjecture: Is the first digit law ubiquitous in the mammalian diaphragm? To address this issue it is important mentioning the necessity to verify if MEPP intervals obey the BL and gBL for other physiological parameters as, for instance, different values for pH and temperature. Indeed, many reports documented that electrical activity of NMJ is modulated by thermal changes \cite{muniak1982}. A remarkable effect of thermal elevation is the acceleration of the vesicular fusion rate, directly reflecting in the MEPP frequency increment. Thus, this experimental maneuver constitutes a propitious inspection of how temperature may affects the degree of conformity. 
	
		Another interesting observation is the relationship between BL/gBL with nonextensive statistics \cite{shao}. Simulations carried out in astrophysical sources and studies done at the NMJ support at least an indirect relation \cite{moret}. Curiously, Pietronero calculations also offer a glimpse toward a formal mathematical connection between gBL and nonextensive theory. Relative to NMJ, Silva \textit{et al.} showed that MEPP histograms are better understood when adjusted with long tails functions, e.g., q-Gaussian distribution \cite{adjesbr}. A nonextensive behavior gives additional support for long-range correlation mechanisms playing role in the communication between nerve and muscle. Furthermore, application of Detrended Fluctuation Analysis (DFA) also uncovered long-range correlations in the MEPP intervals. Since DFA allows detects of scale-invariance in nonstationary data, the findings here reported converge with these previous studies. Thus, the results in this work support the previous theoretical studies in Neuroscience, highlighting a possibility for expanding this study including distributions governed by a lognormal dependence. This function has been applied in electrophysiological signals harvested from NMJ and brain synapses \cite{kloot,buzsaki}. In a nonextensive context, the lognormal distribution has been generalized, taking into account long-range correlations \cite{queiros}. Thus, a simultaneous application of gBL and generalized lognormal function in MEPP intervals, could gives additional evidence toward a common scenario between gBL and nonextensive statistics. The present results attested that gBL emerges as a useful form to identify scale-invariance from electrophysiological time series by noting the evolution of the first digit behavior. Relationships that depending on scale-invariance can have profound implication in Physiology, being manifested in the heart, lung and brain \cite{stanley2001,fadel2004,cai2007,gisiger2000}. It is also claimed that BL observation is intimately connected to an underlying chaotic dynamical process \cite{tolle2000}. In this context, gBL results give a supplementary support to the understanding of scale-invariant properties previously showed at NMJ by other authors \cite{lowen}. In this sense, how such exocytotic machinery can organize themselves into an invariant scale scenario will be the focus of further research. 
	
		Bormashenko asked why BL is frequently observed in statistical data \cite{Bormashenko2016}. According to his view, like BL, many systems entropically governed are described by a logarithm dependence. This argument may explain why BL is so observed in different systems and conditions, including the results described in this work. It is worth mentioning that experiments, carried out at amphibian NMJ, also reported the intervals distribution described by a logarithm behavior \cite{takeda}. Lemons conjectured another intriguing questioning, asking why there are more small things in the world than large things \cite{lemons1985}. In keeping with this author one can paraphrase: Why short MEPP interval, given by the abundance of first digit, prevails over the others ones? Takeda \textit{et al.} had already shown that most intervals concentrate at a shorter time \cite{takeda}. In a physiological interpretation, such short MEPP intervals may be necessary for keeping the spontaneous synaptic activity at a more perennial or secure levels. 
		
	 An important concern in Biophysics is the extrapolation of results from \textit{in vitro} procedure to \textit{in vivo} conditions where tissues are intact. Among disadvantages, the \textit{in vitro} technique presents mechanical stress, promoted during tissue remotion, and artifacts arisen from chemical changes induced by the artificial physiological fluid \cite{azzarelli}. These factors may mask the true electrical response of the neuronal dynamics. On the other hand, \textit{in vitro} methods have many advantages, allowing a better control of several physiological variables, which are not possible using \textit{in vivo} techniques. For example, isolating the neuronal cell enables elimination of efferent contribution responsible for modifying its intrinsic electrical response. Undoubtedly, as any experimental procedure, despite some limitations, \textit{in vitro} techniques are consolidated as a reliable methodology, responsible for innumerable advances of Cell Biophysics. Future studies is required to verify if the BL/gBL may be observed in NMJ of non-mammalians species and pathological tissues \cite {vincent,lewis,shinozaki}. It is also interesting investigating how toxins and drugs could change the first digit distribution in the MEPP intervals. 

\section{Concluding remarks}

In summary, to the best of our knowledge, this is the first report to identify the validity of anomalous numbers law in synaptic transmission. In fact, all experimental data showed conformity with the gBL no matter the statistical method adopted. The results confirm NMJ not only as a remarkable physiological preparation in Biophysics research, but also representing an approach for teaching different topics in the scope of complex systems. In this context, the time series intervals of spontaneous potentials taken from NMJ diaphragm of mice were examined, where BL and gBL validation were robustly observed independently of $[Ca^{2+}]_o$ and $[K^{+}]_{o}$ manipulations. Thus, the present findings convinced us that the spontaneous quantal release conforms the anomalous number phenomenon. Last, but not least, we hope that the present work will serve to motivate the lecturer not only to include the first digit law in their courses. It is also important to propose classroom activities using the first digit law in other biophysical data obtained from repositories.

\ack

We thank prof. Dr. D. Santos and C. Garcia for careful reading of our manuscript. 

\newpage

\appendix

\section*{Appendix}

\setcounter{section}{1}

\begin{table}[!htb]
\caption{MAD and SSD values using BL.} \label{tab3}
\begin{indented}
\item[]\hspace{-2.0cm}\begin{tabular}{@{}ccccccc}
\br
Data  &  1  &  2  &  3  &  4  &  5  &  6  \\ \mr
   &  $ 0.011 $  &  $ 0.009 $  &  $ 0.019 $  &  $ 0.010 $  &  $ 0.008 $  &  $ 0.008 $  \\ 
MAD  &   acceptable   &   acceptable   &   non-   &   acceptable   &   acceptable   &   acceptable   \\ 
   &   conformity   &   conformity   &   conforming   &   conformity   &   conformity   &   conformity   \\ \mr 
   &  $ 25.283 $  &  $ 11.748 $  &  $ 69.49 $  &  $ 20.831 $  &  $ 7.240 $  &  $ 12.285 $  \\ 
SSD  &   marginal   &   acceptable   &   marginal   &   acceptable   &   acceptable   &   acceptable   \\ 
   &   conformity   &   conformity   &   conformity   &   conformity   &   conformity   &   conformity  \\ \mr
\end{tabular}

\item[]\hspace{-2.0cm}\begin{tabular}{@{}ccccccc}
\mr
Data  &  7  &  8  &  9  &  10  &  11  &  12  \\ \mr
   &  $ 0.009 $  &  $ 0.011 $  &  $ 0.012 $  &  $ 0.011 $  &  $ 0.009 $  &  $ 0.011 $  \\ 
MAD  &   acceptable   &   acceptable   &   acceptable   &   acceptable   &   acceptable   &   acceptable   \\ 
   &   conformity   &   conformity   &   conformity   &   conformity   &   conformity   &   conformity   \\ \mr 
   &  $ 10.232 $  &  $ 17.011 $  &  $ 26.874 $  &  $ 13.062 $  &  $ 12.087 $  &  $ 24.118 $  \\ 
SSD  &   acceptable   &   acceptable   &   marginal   &   acceptable   &   acceptable   &   acceptable   \\ 
   &   conformity   &   conformity   &   conformity   &   conformity   &   conformity   &   conformity  \\ \mr
\end{tabular}

\item[]\hspace{-2.0cm}\begin{tabular}{@{}ccccccc}
\mr
Data  &  13  &  14  &  15  &  16  &  17  &  18  \\ \mr
   &  $ 0.020 $  &  $ 0.012 $  &  $ 0.010 $  &  $ 0.009 $  &  $ 0.008 $  &  $ 0.007 $  \\ 
MAD  &   non-   &   acceptable   &   acceptable   &   acceptable   &   acceptable   &   acceptable   \\ 
   &   conforming   &   conformity   &   conformity   &   conformity   &   conformity   &   conformity   \\ \mr 
   &  $ 86.453 $  &  $ 15.426 $  &  $ 16.562 $  &  $ 10.086 $  &  $ 9.113 $  &  $ 7.343 $  \\ 
SSD  &   marginal   &   acceptable   &   acceptable   &   acceptable   &   acceptable   &   acceptable   \\ 
   &   conformity   &   conformity   &   conformity   &   conformity   &   conformity   &   conformity  \\ \mr
\end{tabular}

\item[]\hspace{-2.0cm}\begin{tabular}{@{}ccccccc}
\mr
Data  &  19  &  20  &  21  &  22  &  23  &  24  \\ \mr
   &  $ 0.012 $  &  $ 0.012 $  &  $ 0.013 $  &  $ 0.010 $  &  $ 0.027 $  &  $ 0.024 $  \\ 
MAD  &   acceptable   &   acceptable   &   marginal   &   acceptable   &   non-   &   non-   \\ 
   &   conformity   &   conformity   &   conformity   &   conformity   &   conforming   &   conforming   \\ \mr 
   &  $ 22.422 $  &  $ 21.618 $  &  $ 21.548 $  &  $ 9.911 $  &  $ 117.777 $  &  $ 102.727 $  \\ 
SSD  &   acceptable   &   acceptable   &   acceptable   &   acceptable   &   non-   &   non-   \\ 
   &   conformity   &   conformity   &   conformity   &   conformity   &   conforming   &   conforming  \\ \mr
\end{tabular}

\item[]\hspace{-2.0cm}\begin{tabular}{@{}ccccccc}
\mr
Data  &  25  &  26  &  27  &  28  &  29  &  30  \\ \mr
   &  $ 0.014 $  &  $ 0.016 $  &  $ 0.021 $  &  $ 0.019 $  &  $ 0.019 $  &  $ 0.013 $  \\ 
MAD  &   marginal   &   non-   &   non-   &   non-   &   non-   &   marginal   \\ 
   &   conformity   &   conforming   &   conforming   &   conforming   &   conforming   &   conformity   \\ \mr 
   &  $ 26.107 $  &  $ 33.57 $  &  $ 59.946 $  &  $ 71.313 $  &  $ 55.625 $  &  $ 20.256 $  \\ 
SSD  &   marginal   &   marginal   &   marginal   &   marginal   &   marginal   &   acceptable   \\ 
   &   conformity   &   conformity   &   conformity   &   conformity   &   conformity   &   conformity  \\ \mr
\end{tabular}

\item[]\hspace{0.7cm}\begin{tabular}{@{}cccc}
\mr
Data  &  31  &  32  &  33  \\ \mr
   &  $ 0.014 $  &  $ 0.009 $  &  $ 0.018 $  \\ 
MAD  &   marginal   &   acceptable   &   non-   \\ 
   &   conformity   &   conformity   &   conforming   \\ \mr 
   &  $ 36.178 $  &  $ 11.411 $  &  $ 69.913 $  \\ 
SSD  &   marginal   &   acceptable   &   marginal   \\ 
   &   conformity   &   conformity   &   conformity  \\ \mr
%\br
\end{tabular}

\vspace{-1.0cm}
\end{indented}
\end{table}

\begin{table}[!h]
\caption{Values of $\alpha$, MAD and SSD using gBL.} \label{tab4}
\begin{indented}
\item[]\hspace{-2.0cm}\begin{tabular}{@{}ccccccc}
\br
Data  &  1  &  2  &  3  &  4  &  5  &  6  \\ \mr
$\alpha$  &  $ 0.894 $  &  $ 1.091 $  &  $ 1.249 $  &  $ 1.157 $  &  $ 1.005 $  &  $ 1.097 $  \\ \mr
   &  $ 0.011 $  &  $ 0.006 $  &  $ 0.014 $  &  $ 0.004 $  &  $ 0.008 $  &  $ 0.006 $  \\ 
MAD  &   acceptable   &   close   &   marginal   &   close   &   acceptable   &   close   \\ 
   &   conformity   &   conformity   &   conformity   &   conformity   &   conformity   &   conformity   \\ \mr 
   &  $ 16.777 $  &  $ 5.462 $  &  $ 25.506 $  &  $ 2.386 $  &  $ 7.222 $  &  $ 5.358 $  \\ 
SSD  &   acceptable   &   acceptable   &   marginal   &   acceptable   &   acceptable   &   acceptable   \\ 
   &   conformity   &   conformity   &   conformity   &   conformity   &   conformity   &   conformity  \\ \mr
\end{tabular}

\item[]\hspace{-2.0cm}\begin{tabular}{@{}ccccccc}
\mr
Data  &  7  &  8  &  9  &  10  &  11  &  12  \\ \mr
$\alpha$  &  $ 1.056 $  &  $ 1.100 $  &  $ 1.153 $  &  $ 1.056 $  &  $ 1.012 $  &  $ 0.902 $  \\ \mr
   &  $ 0.008 $  &  $ 0.009 $  &  $ 0.009 $  &  $ 0.010 $  &  $ 0.01 $  &  $ 0.011 $  \\ 
MAD  &   acceptable   &   acceptable   &   acceptable   &   acceptable   &   acceptable   &   acceptable   \\ 
   &   conformity   &   conformity   &   conformity   &   conformity   &   conformity   &   conformity   \\ \mr 
   &  $ 7.880 $  &  $ 9.861 $  &  $ 9.874 $  &  $ 10.672 $  &  $ 11.983 $  &  $ 16.802 $  \\ 
SSD  &   acceptable   &   acceptable   &   acceptable   &   acceptable   &   acceptable   &   acceptable   \\ 
   &   conformity   &   conformity   &   conformity   &   conformity   &   conformity   &   conformity  \\ \mr
\end{tabular}

\item[]\hspace{-2.0cm}\begin{tabular}{@{}ccccccc}
\mr
Data  &  13  &  14  &  15  &  16  &  17  &  18  \\ \mr
$\alpha$  &  $ 1.305 $  &  $ 1.030 $  &  $ 0.858 $  &  $ 0.987 $  &  $ 0.950 $  &  $ 1.028 $  \\ \mr
   &  $ 0.011 $  &  $ 0.011 $  &  $ 0.003 $  &  $ 0.008 $  &  $ 0.007 $  &  $ 0.007 $  \\ 
MAD  &   acceptable   &   acceptable   &   close   &   acceptable   &   acceptable   &   acceptable   \\ 
   &   conformity   &   conformity   &   conformity   &   conformity   &   conformity   &   conformity   \\ \mr 
   &  $ 16.907 $  &  $ 14.812 $  &  $ 1.972 $  &  $ 9.964 $  &  $ 7.333 $  &  $ 6.770 $  \\ 
SSD  &   acceptable   &   acceptable   &   perfect   &   acceptable   &   acceptable   &   acceptable   \\ 
   &   conformity   &   conformity   &   conformity   &   conformity   &   conformity   &   conformity  \\ \mr
\end{tabular}

\item[]\hspace{-2.0cm}\begin{tabular}{@{}ccccccc}
\mr
Data  &  19  &  20  &  21  &  22  &  23  &  24  \\ \mr
$\alpha$  &  $ 0.986 $  &  $ 1.135 $  &  $ 0.873 $  &  $ 1.072 $  &  $ 0.686 $  &  $ 0.734 $  \\ \mr
   &  $ 0.012 $  &  $ 0.008 $  &  $ 0.009 $  &  $ 0.007 $  &  $ 0.020 $  &  $ 0.021 $  \\ 
MAD  &   acceptable   &   acceptable   &   acceptable   &   acceptable   &   non-   &   non-   \\ 
   &   conformity   &   conformity   &   conformity   &   conformity   &   conforming   &   conforming   \\ \mr 
   &  $ 22.263 $  &  $ 7.807 $  &  $ 9.876 $  &  $ 5.968 $  &  $ 45.097 $  &  $ 47.660 $  \\ 
SSD  &   acceptable   &   acceptable   &   acceptable   &   acceptable   &   marginal   &   marginal   \\ 
   &   conformity   &   conformity   &   conformity   &   conformity   &   conformity   &   conformity  \\ \mr
\end{tabular}

\item[]\hspace{-2.0cm}\begin{tabular}{@{}ccccccc}
\mr
Data  &  25  &  26  &  27  &  28  &  29  &  30  \\ \mr
$\alpha$  &  $ 0.873 $  &  $ 0.843 $  &  $ 0.922 $  &  $ 0.753 $  &  $ 0.912 $  &  $ 0.963 $  \\ \mr
   &  $ 0.011 $  &  $ 0.012 $  &  $ 0.020 $  &  $ 0.016 $  &  $ 0.018 $  &  $ 0.011 $  \\ 
MAD  &   acceptable   &   acceptable   &   non-   &   non-   &   non-   &   acceptable   \\ 
   &   conformity   &   conformity   &   conforming   &   conforming   &   conforming   &   conformity   \\ \mr 
   &  $ 14 $  &  $ 16.372 $  &  $ 55.081 $  &  $ 26.039 $  &  $ 50.136 $  &  $ 19.217 $  \\ 
SSD  &   acceptable   &   acceptable   &   marginal   &   marginal   &   marginal   &   acceptable   \\ 
   &   conformity   &   conformity   &   conformity   &   conformity   &   conformity   &   conformity  \\ \mr
\end{tabular}

\item[]\hspace{0.7cm}\begin{tabular}{@{}cccc}
\mr
Data  &  31  &  32  &  33  \\ \mr
$\alpha$  &  $ 0.831 $  &  $ 1.105 $  &  $ 0.753 $  \\ \mr
   &  $ 0.012 $  &  $ 0.004 $  &  $ 0.015 $  \\ 
MAD  &   acceptable   &   close   &   marginal   \\ 
   &   conformity   &   conformity   &   conformity   \\ \mr 
   &  $ 14.640 $  &  $ 3.052 $  &  $ 23.739 $  \\ 
SSD  &   acceptable   &   acceptable   &   acceptable   \\ 
   &   conformity   &   conformity   &   conformity  \\ \mr
\end{tabular}

\vspace{-4.0cm}

\end{indented}
\end{table}

\newpage

\section*{1. Supplementary Material}

\begin{verbatim}
 
##############################################################################
################################						Benford's Law						###############################
##############################################################################


#The code, used in this article, was created by the authors.
#R-cran repository provides two packages for analysis of Benford's Law: 
#benford.analysis and BenfordTests.
#These packages allow the user easily calculate various parameters and perform 
#multiple tests on the data set.
#They can be installed using the command: 
#install.packages("benford.analysis")  # Or Downloaded on the R-cran Repository:
#https://cran.r-project.org/web/packages/benford.analysis/index.html
#install.packages("BenfordTests")      # Or Downloaded on the R-cran Repository:
#https://cran.r-project.org/web/packages/BenfordTests/index.html
#R language reads many formats, we save the input data in the .csv extension.
#Read data by using the "," as decimal point.
data.bf<- read.csv(file = "NMJ-2.csv",header = TRUE,dec = ",")  




###################################################################################

#########
################				Using the Packages to Data Analyses
#########


###		Evaluate the Parameters Using the Package: benford.analysis

library(benford.analysis)		# Activate the package: benford.analysis; that allow you:

model.ba<-benford(data.bf[,1],number.of.digits = 1) # Take the first digit from data
plot(model.ba)                                      # Plot Digits Distribution.

model.ba			      # All information from data.
chisq(model.ba)			# Calculate the Chi-Square test. 
MAD(model.ba)			  # Calculate the Mean Absolute Deviation (MAD) test.



###		Evaluate the Parameters Using the Package: benford.analysis

library(BenfordTests)		# Activate the package: BenfordTests. 
                                       # This package allow you:
model.bt<-data.bf[!is.na( data.bf[,1]),1]  # Remove the NA values
chisq.benftest(model.bt)               # Calculete the Chi-Square test.
ks.benftest(model.bt)                  # Kolmogorov-Smirnov test.
usq.benftest(model.bt)                 # Freedman-Watson U-square test.
mdist.benftest(model.bt)               # Chebyshev Distance test.
edist.benftest(model.bt)               # Euclidean Distance test.
meandigit.benftest(model.bt)           # Judge-Schechter Mean Deviation test.
jpsq.benftest(model.bt)                # Joenssen’s JP-square test.
jointdigit.benftest(model.bt)          # A Hotelling T-square type test.




###################################################################################

#########
################				Parameters Calculation
#########


###		Probability of Benford's Law.
lb<-vector(mode = "numeric"); for(i in 1:9){ lb[i]<-log10(1+1/i ) }


###		First Digit of Data.
first.digit<-data.bf  
for(i in 1:ncol(data.bf)){ first.digit[,i]<-as.numeric( substr(data.bf[,i], 1,1) ) } 


###		Absolute and Relative Frequency of Data.
freq<-first.digit[1,] ; for(i in 1:9){ for( j in 1:ncol(first.digit)){ 
    freq[i,j]<-sum(first.digit[,j]==i,na.rm = T)}}		# Absolute Frequency.                 	

freq.rel<-freq[,c(1,2)] ; for( j in 1:ncol(first.digit)){		# Relative Frequency.
    freq.rel[,j]<- freq[,j]/(sum(freq[,j],na.rm = T) ) } 


###		Calculate Chi-square explicitly.
ep.qui<- colSums( ((freq.rel-lb)^(2)) /lb )* colSums(freq) ; ep.qui


###		Calculate MAD explicitly.
ep.mad<-colSums( abs( freq.rel-lb ))/9 ; ep.mad


###		Calculate SSD explicitly.
ep.ssd<-colSums( ( freq.rel-lb )^(2) )*10^(4) ; ep.ssd


###		Plot Benford's Law Data
barplot( t(data.frame(freq.rel[,1],lb)), main = "Benford's Law Analysis", 
	    	    names.arg =1:9, beside = T, col= c("black","blue"), 
	    	    legend.text = c("Data","BL"), axis.lty = 1,las=1)



###################################################################################

#########
################				Generalized Benford's Law for alpha
#########


###		First Digit of Data.
first.digit<-data.bf 
for(i in 1:ncol(data.bf)){ first.digit[,i]<-as.numeric( substr(data.bf[,i], 1,1) ) } 


###		Absolute and Relative Frequency of Data.
freq<-first.digit[1,] ; for(i in 1:9){ for( j in 1:ncol(first.digit)){
              freq[i,j]<-sum(first.digit[,j]==i,na.rm = T)}}		# Absolute Frequency.             

freq.rel<-freq[,c(1,2)] ; for( j in 1:ncol(first.digit)){		# Relative Frequency.
              freq.rel[,j]<- freq[,j]/(sum(freq[,j],na.rm = T) ) } 


###		Function - Generalized Benford's Law.
prob.alpha<-function(alpha){ aux<- c( (2^(1-alpha) -1)/(1-alpha), (3^(1-alpha) 
      -2^(1-alpha))/(1-alpha), (4^(1-alpha) -3^(1-alpha))/(1-alpha), (5^(1-alpha) 
      -4^(1-alpha))/(1-alpha), (6^(1-alpha) -5^(1-alpha))/(1-alpha), (7^(1-alpha) 
      -6^(1-alpha))/(1-alpha), (8^(1-alpha) -7^(1-alpha))/(1-alpha), (9^(1-alpha) 
      -8^(1-alpha))/(1-alpha), (10^(1-alpha) -9^(1-alpha))/(1-alpha) )
aux<-aux/sum(aux) ; return(aux)}


###		Function - Mean Square deviation for Alpha Variation
alpha.erro<-function(alpha){
  prob.alpha<-c ( (2^(1-alpha) -1)/(1-alpha), (3^(1-alpha) -2^(1-alpha))/(1-alpha),  
        (4^(1-alpha) -3^(1-alpha))/(1-alpha), (5^(1-alpha) -4^(1-alpha))/(1-alpha),  
        (6^(1-alpha) -5^(1-alpha))/(1-alpha), (7^(1-alpha) -6^(1-alpha))/(1-alpha),  
        (8^(1-alpha) -7^(1-alpha))/(1-alpha), (9^(1-alpha) -8^(1-alpha))/(1-alpha),  
        (10^(1-alpha) -9^(1-alpha))/(1-alpha) )
  prob.alpha<-prob.alpha/sum(prob.alpha)
  result<- sum( (prob.alpha-freq.rel[,kk])^2) ; return( result) }


###		To Optimize the Parameter, we use the package: DEoptim. 
#They can be installed using the command: 
#install.packages("DEoptim")  # Or Downloaded on the R-cran Repository: 
#https://cran.r-project.org/web/packages/DEoptim/index.html
library(DEoptim) # Activate the package: DEoptim. This package allow you


###		Optimize the parameter Alpha.
kk<-1  # The column number that will be
alpha.de<-DEoptim(fn=alpha.erro, lower=c(0), upper=c(10),control=list(
	                       NP=200,itermax=1000,strategy=3))   #  Optimize
alpha.de$optim$bestmem[[1]]                #  Best Value for Alpha 
alpha.de$optim$bestval                     #  Error
prob.alpha(alpha.de$optim$bestmem[[1]])    #  Probability of Generalized Benford's Law


###		Chi-Square - Generalized Benford's Law
mep.qui<- colSums(  ((freq.rel[kk]-prob.alpha(alpha.de$optim$bestmem[[1]])  
      )^(2)) / prob.alpha(alpha.de$optim$bestmem[[1]])  )* colSums(freq[kk]) 
mep.qui


###		MAD - Generalized Benford's Law
mep.mad<-colSums( abs( freq.rel[kk]- prob.alpha(alpha.de$optim$bestmem[[1]]) ))/9  
mep.mad 


###		SSD - Generalized Benford's Law
mep.ssd<-colSums( ( freq.rel[kk]- prob.alpha(alpha.de$optim$bestmem[[1]]) 
                     )^(2) )*10^(4)  
mep.ssd



###		Plot Generalized Benford's Law Analysis Data
barplot( t(data.frame(freq.rel[,kk], prob.alpha(alpha.de$optim$bestmem[[1]]) )),  
            main = "Generalized Benford's Law Analysis", names.arg =1:9, beside = T, 
            col= c("black","red"), legend.text = c("Data","gBL"),
            axis.lty = 1,las=1)



\end{verbatim}


\newpage

\section*{References}

	
\begin{thebibliography}{99}


\bibitem{islami}{Islami A and Longo G 2017 {\it Prog. Biophys. Mol. Biol.} \href{https://www.sciencedirect.com/science/article/pii/S0079610717301542?via\%3Dihub}{ {\bf 131} 179} } % Marriages of mathematics and physics: A challenge for biology.

\bibitem{newcomb}{Newcomb S 1881 {\it Am. J. Math.} \href{http://www.jstor.org/stable/2369148}{ {\bf 4} 39}  } % Note of the Frequency of Use of the Different Digits in Natural Numbers.

\bibitem{benford} {Benford F 1938 {\it Proc. Am. Philos. Soc.} \href{http://www.jstor.org/stable/984802}{ {\bf 78 } 551}}% The Law of Anomalous Numbers.

\bibitem{pinkham}{Pinkham R S 1961 {\it  ‎Ann. Math. Stat.} \href{http://www.jstor.org/stable/2237922}{ {\bf 32} 1223}}% On the Distribution of First Significant Digits.

\bibitem{hill95}{Hill T P 1995 {\it Statistical Science} \href{http://www.jstor.org/stable/2246134}{ {\bf 10} 354}} % A Statistical Derivation of the Significant-Digit Law.

\bibitem{pietronero}{Pietronero L, Tosatti E, Tosatti V and Vespignani 2001 {\it Physica A} \href{https://www.sciencedirect.com/science/article/pii/S0378437100006336}{ {\bf 293} 297} } % Explaining the uneven distribution of numbers in nature: The laws of Benford and Zipf.

\bibitem{fu2007}{Fu D, Shi Y Q and Su W 2007 {\it Proc. SPIE } \href{https://www.spiedigitallibrary.org/conference-proceedings-of-spie/6505/65051L/A-generalized-Benfords-law-for-JPEG-coefficients-and-its-applications/10.1117/12.704723.short?SSO=1}{ {\bf 6505} 1} } % A generalized Benford's law for JPEG coefficients and its applications in image forensics. 

\bibitem{gauvrit2009}{Gauvrit N G, Houillon J C and Delahaye J P 2017 {\it Adv. Cogn. Psychol} \href{https://www.ncbi.nlm.nih.gov/pmc/articles/PMC5504535/}{ {\bf 13} 121} } %Generalized Benford’s Law as a Lie Detector.

\bibitem{burke}{Burke J and Kincanon E 1991  {\it Am. J. Phys.} \href{https://aapt.scitation.org/doi/10.1119/1.16838}{ {\bf 59} 952}} % Benford’s law and physical constants: The distribution of initial digits

\bibitem{costas}{Costas E, Lopez-Rodas V, Toro F J and Flores-Moya A 2008 {\it Aquatic Botany} \href{https://www.sciencedirect.com/science/article/pii/S0304377008000533}{ {\bf 89} 341} } % The number of cells in colonies of the cyanobacterium Microcystis aeruginosa satisfies Benford’s law

\bibitem{buck}{Buck B, Merchant A C and Perez S M  1993 {\it Eur. J. Phys.} \href{http://iopscience.iop.org/article/10.1088/0143-0807/14/2/003/meta}{ {\bf 14} 59} } % An illustration of Benford's first digit law using alpha decay half lives

\bibitem{nigrini}{Nigrini M J 2012 {\it Benford's Law: Applications for forensic accounting, auditing, and fraud  detection} (Hoboken: John Wiley \& Sons)}

\bibitem{sinha2015}{Seenivasan P, Easwaran S, Sridhar S and  Sinha S 2016 {\it Front. Physiol.} \href{https://www.frontiersin.org/articles/10.3389/fphys.2015.00390/full}{ {\bf 6} 390} } % Using Skewness and the First-Digit Phenomenon to Identify Dynamical Transitions in Cardiac Models

\bibitem{kreuzer2014}{Kreuzer M, Jordan D, Antkowiak B, Drexler B, Kochs E F and Schneider G 2014 {\it Anesth. Analg.} \href{https://insights.ovid.com/crossref?an=00000539-201401000-00021}{ {\bf 118} 183} } % Brain electrical activity obeys Benford's law.

\bibitem{sudhof}{S\"udhof T C 2013 {\it Neuron} \href{http://www.cell.com/neuron/fulltext/S0896-6273(13)00926-4?_returnURL=https\%3A\%2F\%2Flinkinghub.elsevier.com\%2Fretrieve\%2Fpii\%2FS0896627313009264\%3Fshowall\%3Dtrue}{ {\bf 80} 675}} % Neurotransmitter release: the last millisecond in the life of a synaptic vesicle.

\bibitem{purves}{Purves D, Augustine G J, Fitzpatrick D, Hall W C, LaMantia A S and White L E  2012  {\it Neuroscience} 5 ed (Sunderland: Sinauer Associates)}

\bibitem{sanes}{Sanes J R  and Lichtman J W 1999 {\it Annu. Rev. Neurosci.} \href{https://www.annualreviews.org/doi/10.1146/annurev.neuro.22.1.389}{ {\bf 22} 389} }% DEVELOPMENT OF THEVERTEBRATE NEUROMUSCULAR JUNCTION.

\bibitem{katz1}{Katz B 2003 {\it J. Neurocytol.} \href{https://link.springer.com/article/10.1023\%2FB\%3ANEUR.0000020603.84188.03}{ {\bf 32} 437} }% Neural transmitter release: From quantal secretion to exocytosis and beyond.

\bibitem{katz2}{Fatt P and Katz B 1952 {\it J. Physiol.} \href{https://physoc.onlinelibrary.wiley.com/doi/abs/10.1113/jphysiol.1952.sp004735}{ {\bf 117} 109} } % SPONTANEOUS SUBTHRESHOLD ACTIVITY AT MOTOR NERVE ENDINGS.

\bibitem{bennett}{Bennett M R and Kearns J L 2000 {\it Prog. Neurobiol.} \href{https://www.sciencedirect.com/science/article/pii/S0301008299000404}{ {\bf 60} 545}} %  Statistics of transmitter release at nerve terminals.

\bibitem{washio}{Washio H M and Inouye S T 1980 {\it J. Exp. Biol.} \href{http://jeb.biologists.org/content/87/1/195.short}{ {\bf 87} 195}} % The statistical analysis of spontaneous transmitter release at individual junctions on cockroach muscle.

\bibitem{lowen}{Lowen S B, Cash S S, Poo M and Teich M C 1997 {\it J. Neurosci.} \href{http://www.jneurosci.org/content/17/15/5666.short}{ {\bf 17} 5666} } % Quantal Neurotransmitter Secretion Rate Exhibits Fractal Behavior

\bibitem{takeda}{Takeda T, Sakata A and Matsuoka T 1999 {\it Prog. Neuropsychopharmacol. Biol. Psychiatry.} \href{https://www.sciencedirect.com/science/article/pii/S0278584699000500}{ {\bf 23} 1157} } % Fractal dimensions in the occurrence of miniature end-plate potential in a vertebrate neuromuscular junction.

\bibitem{rizolli}{Rizzoli S O and Betz W J 2005 {\it Nat. Rev. Neurosci.} \href{https://www.nature.com/articles/nrn1583}{ {\bf 6} 57}} % Synaptic vesicle pools.

\bibitem{caboco}{Lima R F, Prado V F, Prado M A and Kushmerick C 2010 {\it J. Neurochem.} \href{https://onlinelibrary.wiley.com/doi/full/10.1111/j.1471-4159.2010.06657.x}{ {\bf 113} 943}} % Quantal release of acetylcholine in mice with reduced levels of the vesicular acetylcholine transporter.

\bibitem{rsoftware}{R Development Core Team 2008 {\it R: A language and environment for statistical computing} (Viena: R Foundation for Statistical Computing)} %R: A language and environment for statistical computing.

\bibitem{rusakov2006}{Rusakov D A 2006 {\it J. Neurochem.} \href{http://journals.sagepub.com/doi/10.1177/1073858405284672}{ {\bf 12} 317} } % Ca2+-Dependent Mechanisms of Presynaptic Control at Central Synapses.

\bibitem{grohovaz}{Grohovaz F, Fesce R and Haimann C 1989 {\it Cell Biol. Int.} \href{https://www.sciencedirect.com/science/article/pii/B9780121764609500161}{ {\bf 13} 1085}} % Dual effect of potassium on transmitter exocytosis. 

\bibitem{adrian1956}{Adrian R H 1956 {\it J. Physiol.} \href{https://physoc.onlinelibrary.wiley.com/doi/abs/10.1113/jphysiol.1956.sp005615}{ {\bf 133} 631} } %  The effect of internal and external potassium concentration on the membrane potential of frog muscle.

\bibitem{kossovsky}{Kossovsky A E 2012 \textit{ Benford Law:Theory, the General Law of Relative Quantities, and Forensic Fraud Detection Applications } (Singapore: World Scientific)}

\bibitem{slepkov2015}{Slepkov A D, Ironside K B and DiBattista D 2015 {\it PLoS ONE} \href{http://journals.plos.org/plosone/article?id=10.1371/journal.pone.0117972}{ {\bf 10} e0117972}} %  Benford’s Law: Textbook Exercises and Multiple-Choice Testbanks

\bibitem{halliday2010}{Halliday A C, Devonshire I M, Greenfield S and Dommett E 2010 {\it Adv. Physiol. Educ.} \href{https://www.physiology.org/doi/10.1152/advan.00005.2010}{ {\bf 34} 205} } % Teaching medical students basic neurotransmitter pharmacology using primary research resources.

\bibitem{falces2015}{Rodriguez-Falces J 2015 {\it Adv. Physiol. Educ.} \href{https://www.physiology.org/doi/10.1152/advan.00130.2014}{ {\bf 39} 15} } % Understanding the electrical behavior of the action potential in terms of elementary electrical sources.


\bibitem{falces2013}{Rodriguez-Falces J 2013 {\it Adv. Physiol. Educ.} \href{https://www.physiology.org/doi/10.1152/advan.00064.2013}{ {\bf 37} 327}} % A novel approach to teach the generation of bioelectrical potentials from a descriptive and quantitative perspective

\bibitem{branco}{Branco T and Staras K 2009 {\it Nat. Rev. Neurosci.} \href{https://www.nature.com/articles/nrn2634}{ {\bf 10} 373}} % The probability of neurotransmitter release: variability and feedback control at single synapses.

\bibitem{muniak1982}{Carlson C G, Kriebel M E and Muniak C G 1982 {\it Neuroscience} \href{https://www.sciencedirect.com/science/article/abs/pii/0306452282902135}{ {\bf 7} 2537} } % The effect of temperature on the amplitude distributions of miniature endplate potentials in the mouse diaphragm.

\bibitem{shao}{ Shao L and Ma B Q 2010 {\it Phys. Rev. E} \href{https://journals.aps.org/pre/abstract/10.1103/PhysRevE.82.041110}{ {\bf 82} 041110} } % First-digit law in nonextensive statistics

\bibitem{moret}{Moret M A, Senna V, Pereira M G and Zebende G F 2006 {\it  ‎Int. J. Mod. Phys. C} \href{https://www.worldscientific.com/doi/abs/10.1142/S0129183106010054}{ {\bf 17} 1597} } % NEWCOMB-BENFORD LAW IN ASTROPHYSICAL SOURCES.

\bibitem{adjesbr}{Silva A J, Lima R F and Moret M A 2011 {\it  Phys. Rev. E} \href{https://journals.aps.org/pre/abstract/10.1103/PhysRevE.84.041925}{ {\bf 84} 041925} } % Nonextensivity and selfaffinity in the mammalian neuromuscular junction. 

\bibitem {kloot} {Kloot W V 1989 {\it J. Neurosci. Methods} \href{https://www.sciencedirect.com/science/article/pii/016502708990054X}{ {\bf 27} 81} } % Statistical and graphical methods for testing the hypothesis that quanta are made up of subunits.

\bibitem {buzsaki}{Buzs\'aki G and Mizuseki K 2014 {\it  Nat. Rev. Neurosci.} \href{https://www.nature.com/articles/nrn3687}{ {\bf 15} 264} } % The log-dynamic brain: how skewed distributions affect network operations.

\bibitem{queiros}{Queir\'os S M D  2012 {\it  Phys. A} \href{https://www.sciencedirect.com/science/article/pii/S0378437112000945}{ {\bf 391} 3594} } % On generalisations of the log-Normal distribution by means of a new product definition in the Kapteyn process.

\bibitem{stanley2001}{Bernaola-Galv\'an P, Ivanov P C, Amaral L A and Stanley H E 2001 {\it Phys Rev Lett.} \href{https://journals.aps.org/prl/abstract/10.1103/PhysRevLett.87.168105}{ {\bf 87} 168105} } % Scale invariance in the nonstationarity of human heart rate.

\bibitem{fadel2004}{Fadel P J, Barman S M, Phillips S W and Gebber G L 2004 {\it J. Appl. Physiol.} \href{https://www.physiology.org/doi/10.1152/japplphysiol.00657.2004}{ {\bf 97} 2056} } %Fractal fluctuations in human respiration.

\bibitem{cai2007}{Cai S M, Jiang Z H, Zhou T, Zhou P L, Yang H J and Wang B H 2007 {\it Phys. Rev. E} \href{https://journals.aps.org/pre/abstract/10.1103/PhysRevE.76.061903}{ {\bf 76} 061903} } % Scale invariance of human electroencephalogram signals in sleep.

\bibitem{gisiger2000}{Gisiger T 2001 {\it Biol. Rev.} \href{https://doi.org/10.1017/S1464793101005607}{ {\bf 76} 161} } % Scale invariance in biology: coincidence or footprint of a universal mechanism? 

\bibitem{tolle2000}{Tolle C R, Budzien J L and LaViolette R A 2000 {\it Chaos} \href{https://aip.scitation.org/doi/abs/10.1063/1.166498}{ {\bf 10} 331} } % Do dynamical systems follow Benford's law?

\bibitem{Bormashenko2016}{Bormashenko E, Shulzinger E, Whyman G and  Bormashenko Y 2016 {\it Physica A} \href{https://www.sciencedirect.com/science/article/pii/S0378437115009498}{ {\bf 444} 524} } %Benford’s law, its applicability and breakdown in the IR spectra of polymers.

\bibitem{lemons1985}{Lemons D S 1986 {\it Am. J. Phys.} \href{https://aapt.scitation.org/doi/10.1119/1.14453}{ {\bf 54} 816} } % On the numbers of things and the distribution of first digits.

\bibitem{azzarelli}{Azzarelli R, Oleari R, Lettieri A, Andre V and Cariboni 2017 {\it Brain Sci.} \href{http://www.mdpi.com/2076-3425/7/5/48}{ {\bf 7} 48} } %In Vitro, Ex Vivo and In Vivo Techniques to Study Neuronal Migration in the Developing Cerebral Cortex.

\bibitem{vincent}{Verschuuren J, Strijbos E and Vincent A 2016 {\it Handb. Clin. Neurol.} \href{https://www.sciencedirect.com/science/article/pii/B9780444634320000244}{ {\bf 133} 447} } %Neuromuscular junction disorders

\bibitem{lewis}{Lewis R L and Gutmann L 2004 {\it Semin. Neurol.} \href{https://www.thieme-connect.de/DOI/DOI?10.1055/s-2004-830904}{ {\bf 24} 175} } %Snake Venoms and the Neuromuscular Junction.

\bibitem{shinozaki}{Shinozaki H 1980 {\it Prog. Neurobiol.} \href{https://www.sciencedirect.com/science/article/pii/0301008280900209}{ {\bf 14} 121} } %Snake Venoms and the Neuromuscular Junction.


\end{thebibliography}
\end{document}